\pdfoutput=1
\newcommand*{\ATLASLATEXPATH}{latex/}
\documentclass[UKenglish,texlive=2011,txfonts,cernpreprint]{\ATLASLATEXPATH atlasdoc}

\usepackage[subfigure]{\ATLASLATEXPATH atlaspackage}
\usepackage{\ATLASLATEXPATH atlasbiblatex}

\usepackage{\ATLASLATEXPATH atlascontribute}

\usepackage{\ATLASLATEXPATH atlasphysics}

\usepackage{amsmath}
\usepackage{bm}
\usepackage{multirow}
\usepackage{placeins}
\usepackage{\ATLASLATEXPATH vgamma-defs}
\usepackage{graphicx}
\usepackage{caption}
\usepackage{amssymb}

\AtlasJournalRef{Physics Letters B 764 (2017) 11–30}
\AtlasDOI{10.1016/j.physletb.2016.11.005}

\graphicspath{{logos/}{figures/}}

\AtlasTitle{Search for heavy resonances decaying to a $Z$ boson and a
  photon in $pp$ collisions at $\sqrt{s}=13$~\TeV\ with the ATLAS detector}

\author{The ATLAS Collaboration}

\AtlasRefCode{EXOT-2016-02}

\PreprintIdNumber{CERN-EP-2016-163}

\AtlasDate{\today}

\AtlasJournal{Phys.\ Lett.\ B.}

\AtlasAbstract{This Letter presents a search for new resonances 
with mass larger than 250~\GeV, decaying to a $Z$ boson and a photon.
The dataset consists of an integrated luminosity of 3.2~fb$^{-1}$
of $pp$ collisions collected at $\sqrt{s} = 13$~\TeV\ with
the ATLAS detector at the Large Hadron Collider. 
The $Z$ bosons are identified through their decays either to charged, 
light, lepton pairs ($e^+e^-$, $\mu^+\mu^-$) or to hadrons. 
The data are found to be consistent with the expected background in the whole 
mass range investigated and upper limits are set on the production
cross section times decay branching ratio to $Z\gamma$ of a narrow
scalar boson with mass between 250~\GeV\ and 2.75~\TeV.
}

\hypersetup{pdftitle={ATLAS document},pdfauthor={The ATLAS Collaboration}}

\begin{document}

\maketitle

\section{Introduction}
\label{sec:introduction}

Many models of physics beyond the Standard Model (SM) introduce
new bosons through either an extension of the Higgs sector or
additional gauge fields.  
This suggests that a broad experimental survey of physics beyond the
SM can be made by searching for new massive bosons. Some models
predict that these bosons decay to final states containing the SM
electroweak $W$ or $Z$ bosons or photons~\cite{Theory-LSTC, Low:2011gn}.
Attractive decays from an experimental perspective are to
$\gamma\gamma$~\cite{Aad:2015mna,Khachatryan:2015qba,Aaboud:2016tru,Khachatryan:2016yec},
$Z\gamma$~\cite{Aad:2013izg,Aad:2014fha}
 or $ZZ$~\cite{Khachatryan:2014gha,Aad:2015ipg} final states, since both
the $Z$ bosons and photons in pair production can be measured well with
relatively low backgrounds. 
If such new bosons were produced, the complete reconstruction of these
final states could be used to precisely measure their properties,
such as their mass.

This Letter presents a search for $X\to Z\gamma$ resonances
using an integrated luminosity of 3.2 fb$^{-1}$ of proton--proton
($pp$) collisions at a centre-of-mass energy $\sqrt{s}$ of 13~\TeV,
collected with the ATLAS detector at the Large
Hadron Collider (LHC) in 2015.
To enhance the sensitivity of the search, both the leptonic
($Z\to \ell^+\ell^-$, $\ell=e,\mu$)\footnote{In the following,
  $\ell^+\ell^-$ final states are referred to as $\ell\ell$ for
  simplicity.} and
hadronic ($Z\to q\bar{q}$) decay modes of the $Z$ boson are used.
The combined selection captures about 77\% of all $Z$ boson decays.
In the following, the search
based on the selection of $\ell\ell\gamma$ final states is also
referred to as the {\em leptonic} analysis, while the
search based on the selection of the $q\bar{q}\gamma$ final state
is also referred to as the {\em hadronic} analysis.

The leptonic analysis uses events collected using lepton triggers
and is performed in the $X$ boson mass ($m_X$) range 250~\GeV--1.5~\TeV.
The hadronic analysis is performed in the $m_X$
range 700~\GeV--2.75~\TeV. Due to the large value of $m_X$,
the $Z$ bosons from $X\to Z\gamma$ are highly boosted and the
two collimated sprays of energetic hadrons, called jets in the following,
that are produced in $Z\to q\bar{q}$ decays are merged into a single,
large-radius, jet $J$. The events used for the hadronic analysis are
collected using single-photon triggers.
Due to the larger $Z$ boson branching ratio to hadrons,
the boosted hadronic analysis dominates the sensitivity at high $m_X$,
where the number of events is very small, while the leptonic analysis,
with its higher signal-to-background ratio, dominates the sensitivity at low $m_X$. 

Previous searches for non-SM bosons decaying into $Z\gamma$ final
states were carried out at the Tevatron and the LHC.  
The D0 Collaboration set limits~\cite{Abazov:2008wg} on $X\to$ $Z\gamma$
production using $p\bar{p}$ collisions at $\sqrt{s}$ = 1.96~\TeV.
At the LHC, the ATLAS
Collaboration used $pp$ collisions
collected in 2011 and 2012 at $\sqrt{s}=7$ and 8~\TeV\ to extend the mass
range and sensitivity of $X\to$ $Z\gamma$ searches~\cite{Aad:2013izg,Aad:2014fha}.
The analyses assumed a narrow width for the $X$ boson
and used $e^+e^-$ and $\mu^+\mu^-$ decays of the $Z$ boson. No signals
were observed and limits on the product of the production cross section
$\sigma(pp\to X)$ times the branching ratio $BR(X\to Z\gamma)$
were determined for values of $m_X$ in the range $\approx$ 200 to 1600~\GeV.  
  
The analyses presented here search for a localized
excess in the reconstructed invariant mass distribution of the
final state, either a photon and two leptons or a photon and a heavy,
large-radius jet.
In the leptonic analysis,
the main background arises from continuum production of a $Z$ boson
in association with a photon, or, to a lesser extent, 
with a hadronic jet misidentified as a photon. 
In the hadronic analysis, the background is dominated
by non-resonant SM production of $\gamma+$jet events, with smaller
contributions from dijet events with a jet misidentified as a photon,
and from SM $V+\gamma$ events ($V=W,Z$).
The invariant mass distribution of the background should be smoothly
and steeply decreasing with the mass. It is parameterized
by a smooth function with free parameters, which are adjusted to the data.
The intrinsic width of the heavy boson is assumed to be small
compared to the experimental resolution.
The boson is assumed to be a spin-0 particle produced via gluon
fusion.

\section{The ATLAS detector}
\label{sec:detector}

The ATLAS detector is a multi-purpose particle
detector with approximately forward-backward symmetric
cylindrical geometry.\footnote{ATLAS uses a right-handed coordinate
system with its origin at the nominal interaction point (IP) in the
centre of the detector and the $z$-axis along the beam pipe. The
$x$-axis points from the IP to the centre of the LHC ring, and the
$y$-axis points upward. Cylindrical coordinates $(r,\phi)$ are used
in the transverse plane, $\phi$ being the azimuthal angle around the
$z$-axis. The pseudorapidity is defined in terms of the polar angle
$\theta$ as $\eta=-\ln\tan(\theta/2)$.}
Its original design~\cite{atlas-detector} has been complemented with 
the installation, prior to the 2015 data-taking, of a new
innermost silicon pixel layer~\cite{Capeans:1291633}.

A two-level trigger system~\cite{Aad:2012xs} selects events to be recorded 
for offline analysis. 
The first-level trigger is hardware-based, while the second, 
high-level trigger is implemented in software and employs algorithms 
similar to those used offline to identify lepton and photon candidates.

\section{Data sample}
\label{sec:data}

Data were collected in 2015
during $pp$ collisions at a centre-of-mass energy of 13~\TeV.
The bunch spacing was 25 ns and the average
number of inelastic interactions per bunch crossing was 13.

The search in the $\ell\ell\gamma$ final state is performed in 
events recorded using the lowest-threshold unprescaled single-lepton
or dilepton triggers.
The single-muon trigger has a nominal transverse momentum (\pT) threshold
of 20~\GeV\ and a loose requirement on the track isolation.
This quantity, defined as the sum of the transverse momenta of 
 the tracks in the inner detector (ID) found in a cone 
of size $\Delta R \equiv \sqrt{(\Delta\eta)^{2} + (\Delta\phi)^{2}} = 0.2$
 around the muon,
excluding the muon track itself, is required to be less than 12\% of
the muon $\pT$. Only tracks with longitudinal impact parameter 
$z_0$ within 6~mm of that from the
muon track are considered in the calculation.
An additional single-muon trigger with a higher $\pT$ threshold (50~\GeV) 
but no isolation requirement is also used.
The dimuon trigger has a $\pT$ threshold of 10~\GeV\ for 
both muon candidates and applies no isolation criteria.
The single-electron (dielectron) trigger has a nominal $\pT$
threshold of 24~\GeV\ (12~\GeV).
Electron candidates are required to satisfy likelihood-based
identification criteria looser than those applied offline and
described in Section~\ref{sec:selection}. The electron identification
likelihood is computed from both the properties of the track reconstructed
in the ID and the energy deposited in the
electromagnetic (EM) calorimeter. 

The search in the $J\gamma$ final state uses events
recorded by the lowest-$\pT$ threshold 
unprescaled single-photon trigger.
This trigger requires at least one photon candidate 
with $\pT>120$~\GeV\ passing 
loose identification requirements based on the
shape of the shower in the EM calorimeter and on the
energy leaking into the hadronic calorimeter~\cite{Aad:2010sp}. 

The trigger efficiency for events satisfying the offline
selection criteria described in Section~\ref{sec:selection}
is greater than 99\% in the $ee\gamma$ and $J\gamma$ channels
and is about 96\% in the $\mu\mu\gamma$ channel due to the reduced
geometric acceptance of the muon trigger system.

The integrated luminosity after the trigger and data quality
requirements is $L_\mathrm{int}=3.2$~\ifb.

\section{Monte Carlo simulation}
\label{sec:mc}

Simulated signal and background samples were generated with a 
Monte Carlo (MC) technique.
They are used to optimize the selection criteria and 
to quantify the signal efficiency of the final selection.
Such MC samples are also used to test the analytic parameterization
of the $Z\gamma$ invariant mass spectra of signal and 
background, while the estimate of the background yield
after the selection is estimated {\em in situ} from the data.

All MC samples are generated assuming a centre-of-mass
$pp$ collision energy of 13~\TeV.
The samples are passed through a detailed simulation of the ATLAS
detector response~\cite{Aad:2010ah} based
on~\textsc{Geant4}~\cite{Agostinelli2003250}.
Multiple inelastic proton--proton collisions (referred to as pile-up) are simulated
with the soft QCD processes of \PYTHIA 8.186~\cite{Sjostrand:2007gs} using the A2
set of tuned parameters (A2 tune)~\cite{Atlas:ATL-PHYS-PUB-2012-003}
and the MSTW2008LO parton distribution function (PDF)
set~\cite{Martin:2009iq}, and are overlaid
on each MC event.
The distribution of the number of pile-up interactions in the simulation
is reweighted to match the data.
The simulated signals in the detector are passed through the
event reconstruction algorithms used for the data.
The simulation is tuned to take into account small differences with data.
These include corrections to photon, lepton and jet reconstruction and
selection efficiencies, and their energy or momentum resolution and scale.
The corrections are obtained either from control samples selected in
early $\sqrt{s}=13$~\TeV\ data or from 8~\TeV\ data with additional
systematic uncertainties introduced to cover the different conditions
between the 2012 and 2015 data-taking.

In the signal simulation, a scalar boson $X$ is produced in 
$pp$ collisions via gluon fusion, and 
decays to a photon and a $Z$ boson. 
Monte Carlo samples are produced for different $m_X$ hypotheses 
between 200~\GeV\ and 3~\TeV.
The width of the boson $X$ is set to 4~\MeV,
which is much smaller than the experimental resolution,
regardless of the resonance mass.
Due to the assumed narrow width of the $X$ boson
and the small contribution of gluon fusion
to the non-resonant SM production 
of $Z$+$\gamma$~\cite{Grazzini:2013bna}, the interference between the 
$gg \to X\to Z\gamma$ signal process and the
SM $gg \to Z\gamma$ background is
neglected in the simulation.
The signal samples are generated with
{\normalfont \scshape POWHEG-BOX}~\cite{Alioli:2010xd,Bagnaschi:2011tu} interfaced to
\PYTHIA 8.186 for the underlying event, parton showering and hadronization.
The CT10~\cite{Lai:2010vv} PDF set and
the AZNLO tune~\cite{Aad:2014xaa} of the underlying
event are used.

Events from SM processes containing a photon and a $Z$ or $W$ 
boson ($V+\gamma$),
a $Z$ boson produced in association with jets, or a prompt photon
produced in association with jets ($\gamma+$jets)
are simulated using the \SHERPA
2.1.1~\cite{Gleisberg:2008ta} generator.
The matrix elements for SM $V+\gamma$ ($\gamma +$jets) production
are calculated for real emission of up to three (four) partons
at leading order (LO) in the strong coupling constant 
$\alpha_{\mathrm{S}}$ and are merged with the \SHERPA parton
shower~\cite{Schumann:2007mg} using the ME+PS@LO
prescription~\cite{Hoeche:2009rj}.
The matrix elements of events containing $Z$ bosons with associated
jets are calculated for up to two partons at next-to-leading order (NLO)
and four partons at LO and merged with the parton shower using
the ME+PS@NLO prescription~\cite{Hoeche:2012yf}.
The matrix elements are calculated using the Comix~\cite{Gleisberg:2008fv} and
OpenLoops~\cite{Cascioli:2011va} generators.
For all the background samples, the CT10 PDF set is used in conjunction
with dedicated parton shower tuning developed by the \SHERPA authors.
The $\gamma+$jets and $V+\gamma$ samples are generated in binned ranges
of the transverse momentum of the photon to ensure precise
predictions over the full spectrum relevant for these
analyses. Similarly, $Z+$jets events are generated in binned ranges 
of the dilepton pair $\pT$ from the $Z$ boson decays.

\section{Event selection}
\label{sec:selection}

Events with at least one primary vertex candidate with two or more
tracks with $\pt > 400$~\MeV\ are selected.
In each event, the primary vertex candidate with the largest sum
of the $\pt^2$ of the associated tracks is chosen as the hard
interaction primary vertex.

Events are required to contain at least one photon candidate
and one $Z$ boson candidate.
In the leptonic analysis, the $Z$ boson candidate is
formed from a pair of opposite-sign, same-flavour electrons or muons.
In the hadronic analysis, $Z$ bosons are required to
recoil against a high-momentum photon ($\pT>250$~\GeV);
as a consequence of the $Z$ boson's large Lorentz boost, the two
jets from the hadronization of the two quarks
are reconstructed as a single, relatively heavy,
large-radius jet. Jet-substructure variables
and the jet mass are then used to discriminate
between a $Z$ boson decay and jets from single quarks or
gluons~\cite{ATL-PHYS-PUB-2015-033}.
Events with one or more electron or muon
candidates satisfying the selection described below
are vetoed in the hadronic analysis.
In the following, the selection of photons, leptons,
large-radius jets and of the final $X\to Z\gamma$
candidates is described.

Unconverted photons, photon conversions to electron-positron pairs,
and electrons are reconstructed from clusters of energy deposits
in the EM calorimeter cells found by a sliding-window algorithm
and from tracks reconstructed in the ID and
extrapolated to the calorimeter~\cite{Aaboud:2016yuq,ATLAS-CONF-2014-032}. 

Photon candidates are required to have a pseudorapidity within 
the regions $|\eta|<1.37$ or $1.52<|\eta|<2.37$, where the first 
calorimeter layer has high granularity.
In the leptonic analysis, the transverse momentum of photon
candidates is initially required to pass a loose preselection,
$\pT>15$~\GeV, whereas the final photon $\pT$ requirement is applied when
a $Z\gamma$ candidate is reconstructed, as described later.
In the hadronic analysis, the photon transverse momentum is required
to be larger than 250~\GeV.
To reduce background from hadronic jets, photon candidates are required
to satisfy a set of requirements on the shower leakage in the hadronic
calorimeter and on
the transverse shower profile measured with the first two
layers of the electromagnetic calorimeter~\cite{Aaboud:2016yuq}.
The requirements were optimized
using simulated samples of photons and hadronic jets produced in
13~\TeV\ $pp$ collisions.
The efficiency of the identification criteria is about
98\% for converted photon candidates and 94\% for unconverted
photon candidates with $\pT>100$~\GeV.
Background from hadronic jets is further reduced by requiring
the transverse energy measured in the calorimeter in a cone
of size $\Delta R = 0.4$ around the photon direction
(\etiso~\cite{Aad:2012tba}, also called {\em calorimeter isolation} in the following)
to be less than $2.45\GeV + 0.022\times p_\mathrm{T}$.

Electron candidates are required to have 
$\pT > 10$~\GeV\ and $|\eta| < 2.47$, excluding the
transition region between the barrel and endcaps in the EM 
calorimeter ($1.37 < |\eta| < 1.52$). 
To suppress background from hadronic jets,
electron candidates are required to satisfy likelihood-based
identification criteria~\cite{ATLAS:2016iqc}.
Such requirements provide 
approximately 85\% identification efficiency
for electrons with a transverse momentum of 20~\GeV,
increasing to 95\% for $\pT>80$~\GeV.

Muons with $|\eta|<2.5$ are reconstructed by combining
tracks in the ID with tracks in the muon spectrometer (MS)~\cite{Aad:2016jkr}.
The acceptance is extended to the region $2.5<|\eta|<2.7$ by also selecting
muons whose trajectory is reconstructed only in the MS.
Muon candidates are required to have transverse momentum above
10~\GeV.
Background muons, originating mainly from pion and kaon decays,
are rejected by applying a set of quality requirements on the
number of hits in the muon spectrometer
and (for $|\eta|<2.5$) on the compatibility between the
ID and MS momentum measurements.
The muon identification efficiency is around
97\% for transverse momenta above 10~\GeV.

If two electron candidates share the same track,
or have clusters in the calorimeter separated by $|\Delta\eta| < 0.075$
and $|\Delta\phi| < 0.125$, only the candidate with the higher
energy measured by the calorimeter is kept.
In addition, if the track associated with an electron candidate is within
a distance $\Delta R = 0.02$ from the track associated with a muon candidate,
the electron candidate is rejected.

Track and calorimeter isolation requirements are further applied
to the selected leptons.
For electrons, combined criteria are applied to the calorimeter isolation,
\etiso, in a cone of radius $\Delta R=0.2$, and to the track isolation,
$\sum_\mathrm{tracks} \pT$, in a cone of radius $\Delta R = 0.2$ for 
electron transverse momenta $\pT < 50$~\GeV\
and of radius $\Delta R = (10~\GeV)/\pT$ for $\pT > 50$~\GeV.
In the calculation of the track isolation, the contribution from
the electron track itself is not included. The criteria are chosen to
provide an efficiency of about 99\% independent of the electron
transverse momentum and pseudorapidity, as determined in a control
sample of $Z\to ee$ decays selected with a tag-and-probe
technique~\cite{ATLAS:2016iqc}.
For muons, combined criteria are imposed on \etiso\ in a cone of
radius $\Delta R=0.2$ and on $\sum_\mathrm{tracks}{\pT}$ inside a cone
of radius $\Delta R=0.3$ for muon transverse momenta $\pT<33$~\GeV\
and of radius $\Delta R=(10~\GeV)/\pT$
for $\pT>33$~\GeV. The efficiency of these criteria increases with
the muon transverse momentum, reaching 95\% at 25~\GeV\ and 99\%
at 60~\GeV, as measured in $Z\to\mu\mu$ events selected
with a tag-and-probe method~\cite{Aad:2016jkr}.

In the hadronic analysis, topological clusters of energy in the calorimeter that were
locally calibrated and assumed to be massless~\cite{Aad:2016upy} 
are used as inputs to reconstruct
large-radius jets, based on the  anti-$k_t$ algorithm~\cite{antikt}
with radius parameter $R = 1.0$~\cite{Aad:2015rpa}.
Within the large-radius jets, smaller ``subjets'' are reconstructed using the
$k_\perp$ algorithm~\cite{Ellis_kt,Catani:1993hr} with a radius parameter
$R = R_\mathrm{sub} = 0.2$. The
large-radius jet is trimmed~\cite{Krohn:2009th} by removing subjets
that carry fractional $\pT$
less than $f_\mathrm{cut} = 5$\% of the $\pT$ of the original jet.  
The pseudorapidity, energy and mass of these trimmed large-radius jets are
calibrated using a simulation-based calibration scheme~\cite{Aad:2011he}.  
The large-radius jets are required to have $\pT>200$~\GeV\
and $|\eta|<2.0$.
Large-radius jets within $\Delta R=1.0$ from selected photons are
discarded.
A $\pT$--dependent requirement on the substructure observable
$D_2^{(\beta=1)}$~\cite{Larkoski:2014gra}, 
defined as the ratio $e_3^{(\beta=1)}/\left(e_2^{(\beta=1)}\right)^3$
of $N$-point energy correlation functions $e_N^{(\beta=1)}$ of the jet 
constituents~\cite{Larkoski:2013eya}, is
used to select hadronically decaying bosons while rejecting 
jets from single quarks or gluons. The ratio makes use of the sensitivity
of the $e_N$ functions to the ``pronginess'' character of the jet.
In particular, it relies on the sensitivity of $e_2$ to radiation around
a single hard core, and of $e_3$ to radiation with two cores.
The powers of the $e_2$ and $e_3$ functions in the ratio are chosen to
optimize the discrimination between one- and two-prong jets following
an analysis of the ($e_2$, $e_3$) phase-space of these two types of jets.

The jet mass $m_J$, computed from its topological cluster constituents
that remain after the trimming procedure, is required to be in the
range 80~\GeV$< m_J <110$~\GeV.
The jet is required to be associated with less than 30 tracks 
with $\pT>500$~\MeV\ originating from the hard-interaction
primary vertex (before trimming).
The efficiency of the $D_2^{(\beta=1)}$, $m_J$ and number-of-track
requirements is around 22\% for the signal jet and 2.2\% for
jets from single quarks or gluons.

After the selection of photons, leptons and large-radius jet candidates,
the $Z\gamma$ candidate is chosen.
If an event has multiple photon or jet candidates, only the photon or jet
candidate with highest transverse momentum is kept.
In the leptonic analysis, only $Z\to\ell\ell$ candidates with invariant
mass $m_{\ell\ell}$ within $\pm 15$~\GeV\ of the $Z$ boson 
mass~\cite{Agashe:2014kda} are retained;
in case of multiple dilepton candidates, only the one with invariant mass
closest to the $Z$ boson mass is kept. Moreover,
the triggering leptons are required to match one, or both in the case of
events collected with dilepton triggers, of the $Z$ boson candidate's leptons.

The invariant mass \mrec\ of the selected $Z\gamma$ candidate is 
computed from the four-momenta of the photon candidate
and either the selected leptons or the jet
($\mrec=m_{\ell\ell\gamma}$ or $m_{J\gamma}$). 
In the leptonic analysis, the four-momentum of the photon is
recalculated using the identified primary vertex as the photon's origin,
while the four-momenta of the leptons are first corrected for collinear FSR
(muons only) and then recomputed by means of a
$Z$-mass-constrained kinematic fit~\cite{Aad:2012tfa}.
The $Z\gamma$ invariant mass is required to be larger than 200 (640)~\GeV\
for the leptonic (hadronic) analysis, to be sufficiently far from the kinematic
turn-on due to the $Z$ boson mass and the photon transverse 
momentum requirement.

Finally, the leptonic analysis only retains candidates in which the
photon transverse momentum is larger than 30\% of \mrec,
significantly suppressing background at large invariant mass while 
maintaining high efficiency over a large range of signal masses.

\section{Signal and background models}
\label{sec:models}
The final discrimination between signal and background events
in the selected sample is achieved by means of an unbinned
maximum-likelihood fit of a signal+background model to the
invariant mass distribution of the selected data events.
Both the signal and background models are described in this section.

\subsection{Signal model}
\label{ssec:signal_model}

Figure~\ref{fig:signal_mass} illustrates the distributions of
$m_{\ell\ell\gamma}$ and $m_{J\gamma}$
for simulated signal events for a resonance mass of 800~\GeV.
The intrinsic width of the simulated resonance (4~\MeV) is negligible compared to
the experimental resolution.
The $m_{\ell\ell\gamma}$ resolution
ranges between 2~\GeV\ at $m_X=200$~\GeV\ and 15~\GeV\ 
at $m_X=1500$~\GeV\ (1\% relative resolution).
The $m_{J\gamma}$ resolution ranges between 22~\GeV\ at
$m_X=750$~\GeV\ (3\%) and 50~\GeV\ at $m_X=3$~\TeV\ (1.7\%).

\begin{figure}[h]
  \begin{center}
    \includegraphics[width=0.9\columnwidth]{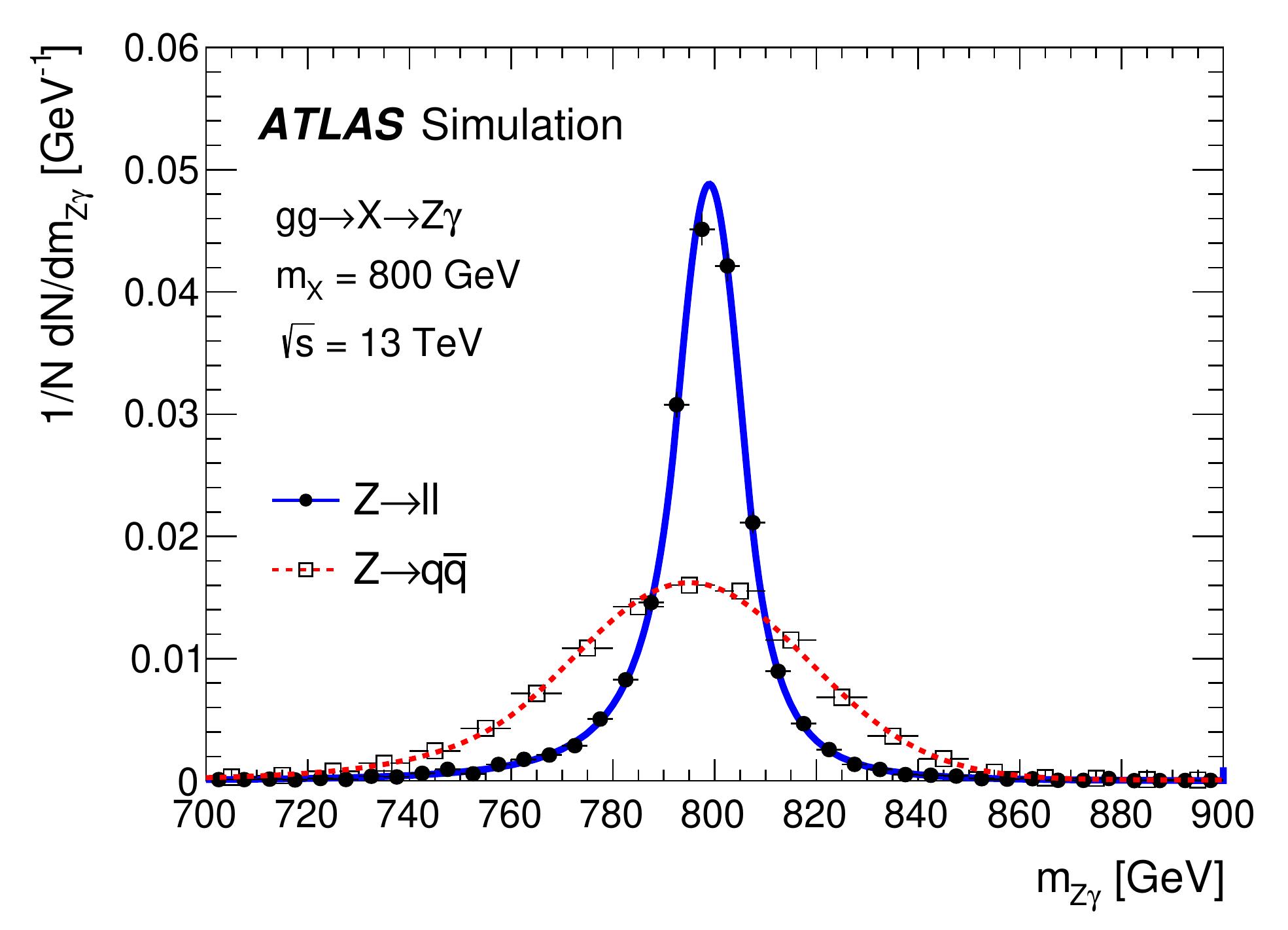}
    \caption{Invariant-mass distribution for $X\to Z\gamma$,
      $Z\to \ell\ell$ (solid circles) or $Z\to q\bar{q}$ events (open squares) in a simulation
      of a narrow resonance $X$ with a mass of 800~\GeV\ produced
      in a gluon-fusion process in $\sqrt{s}=13$~\TeV\ $pp$ collisions.
      All selection requirements have been applied.
      The blue solid (red dashed) line represents the fit of the points with a double-sided
      Crystal Ball function (sum of a Crystal Ball function and a Gaussian function).}
    \label{fig:signal_mass}
  \end{center}
\end{figure}

The $m_{\ell\ell\gamma}$ distribution is modelled with a double-sided 
Crystal Ball function
(a Gaussian function with power-law tails on both sides).
The $m_{J\gamma}$ distribution is modelled with the sum of
a Crystal Ball function~\cite{CrystalBall} 
(a Gaussian function with a power-law tail on one side)
and a second small, wider Gaussian component.
The fraction of signal $J\gamma$ events described by the Crystal Ball function
is above 90\% for resonance masses up to 1.8~\TeV\ and decreases
with $m_X$, reaching 85\% at $m_X=3$~\TeV.
Polynomial parameterizations of the signal shape parameters as a
function of the resonance mass $m_X$ are obtained from a simultaneous fit to
the invariant mass distributions of all the simulated signal samples,
for each $Z$ boson decay channel.

The signal detection efficiency (including the acceptance of the kinematic
criteria) as a function of $m_X$ is computed
in the leptonic analysis by interpolating
the efficiencies predicted by all the simulated signal samples up to 
$m_X=1.5$~\TeV\ with a function of the form $a + b \mathrm{e}^{c m_X}$.
In the hadronic analysis, the efficiency
at any value of $m_X$ is obtained through a 
linear interpolation between the efficiencies obtained
from the two simulated signal samples with masses closest to $m_X$.
The signal detection efficiency of the leptonic analysis
ranges between 28\% at $m_X=250$~\GeV\ and 43\% at $m_X=1.5$~\TeV,
while that of the hadronic analysis increases from
11\% at $m_X = 700$~\GeV\ to 15\% at $m_X=3$~\TeV, as shown in
Figure~\ref{fig:signal_efficiency}.

\begin{figure}[!htbp]
  \begin{center}
    \includegraphics[width=0.9\columnwidth]{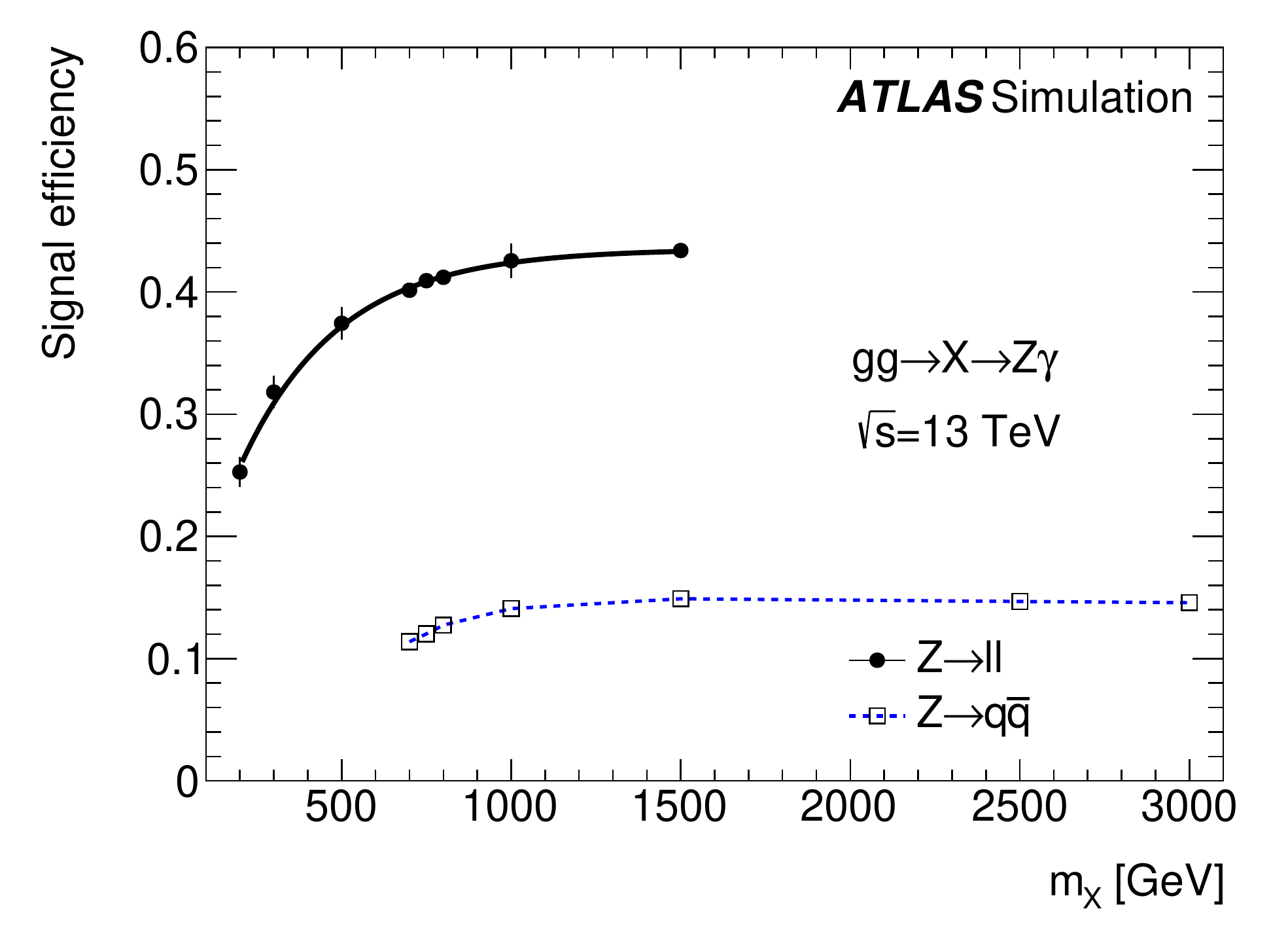}
    \caption{Efficiency (including the acceptance of the kinematic criteria)
      of the leptonic selection for simulated
      signal events in which $Z$ bosons decay to $\ell\ell$ 
      (solid circles),
      and of the hadronic selection for simulated signal events
      in which the $Z$ bosons decay to $q\bar{q}$
      (open squares), as a function of the resonance mass $m_X$.
      The solid line represents an interpolation with a smooth function 
      (of the type $a+b\mathrm{e}^{cm_X}$) of
      the leptonic analysis efficiency, while the dashed line represents
      a linear, piece-wise interpolation of the efficiencies 
      of the hadronic analysis.
    }
    \label{fig:signal_efficiency}
  \end{center}
\end{figure}

\subsection{Background model}
\label{ssec:background_model}

In both the leptonic and hadronic final states, the total background
exhibits a smoothly falling spectrum as a function of the invariant
mass \mrec\ of the final-state products.
The \mrec\ distribution of the background is parameterized with
a function similar to the one 
used in previous searches in the $\gamma+$jet and diphoton
final states~\cite{EXOT-2012-22,Aaboud:2016tru}:
\begin{equation}
 f_\mathrm{bkg}(\mrec) = \mathcal{N} (1-x^k)^{p_1+\xi p_2}x^{p_2}.
 \label{back:eqn:background}
\end{equation}

Here $\mathcal{N}$ is a normalization factor, $x=\mrec/\sqrt{s}$,
the exponent $k$ is $1/3$ for the leptonic analysis and 1
for the hadronic analysis, and $p_1$ and $p_2$ are dimensionless shape
parameters that are fitted to the data. The constant $\xi$
is set to zero in the leptonic analysis and to the value
(ten) that minimizes the correlation between
the maximum-likelihood estimates of $p_1$ and $p_2$ in a fit
to the background simulation for the hadronic analysis.

These parameterizations were chosen since they satisfy
the following two requirements: (i) the bias in the fitted signal
due to the choice of this functional form is estimated to be
sufficiently small compared to the statistical uncertainties from the
background, and (ii) the addition of further degrees of freedom to
Eq.~(\ref{back:eqn:background}) does not lead to a significant
improvement in the goodness of the fit to the data distribution. 

The bias is checked by performing signal+background fits
to large background control samples, scaled to
the luminosity of the data.
A functional form is retained if the absolute value
of the fitted signal yield 
$N_{\mathrm{spur}}$ ({\em spurious signal}
in the following) is less than 20\% (25\%)
of its statistical uncertainty in the leptonic
(hadronic) analysis~\cite{Aad:2014eha}.

For the leptonic analysis, the control sample for the spurious signal study
is obtained by summing the invariant mass distributions of
$Z+\gamma$ and $Z+$jets simulated
events, normalized according to their relative fractions measured in data
(90\% and 10\% respectively).
These fractions are determined by means of a simultaneous fit of the 
$\etiso$ distributions of the photon candidates passing or failing
the identification requirements.
To increase the number of $Z+\gamma$ MC events, a very large (up to one
thousand times more events than in data) simulated sample is obtained by
passing the events generated by \SHERPA through a
fast simulation of the calorimeter response~\cite{ATLAS:1300517}.
The agreement of the \mrec\ distribution in the parametric simulation
with that of the full-simulation $Z+\gamma$ sample described in
Section~\ref{sec:mc} was evaluated with a $\chi^2$ test.
The $\chi^2$ was found to be 23 for 28 degrees of freedom,
corresponding to a $p$-value of 75\%, 
indicating that the shapes agree well within statistical uncertainties.
The \mrec\ distribution of $Z+$jets
events is obtained by reweighting that of the large $Z+\gamma$
sample by a second-order polynomial function. The parameters of this
function are determined from a fit to the ratio of the \mrec\ distributions
of a $Z+$jets-enriched data control sample to that of the parameterized
simulation of $Z+\gamma$.

For the hadronic analysis, the spurious signal is studied in a data
control sample enriched in jets not originating from $Z$ boson decays.
This sample passes the selection described in
Section~\ref{sec:selection}, with the exception that 
the jet mass $m_J$ is either between 50~\GeV\ and 65~\GeV, or between
110~\GeV\ and 140~\GeV. 
Based on simulation and data-driven studies, 
the $m_{J\gamma}$ distribution of  $\gamma+$jets events has a similar
shape to that of the total background in the signal region,
where the latter also includes contributions at the 10\% level from
$V + \gamma$ and dijet events.
Thus, this control region (dominated by $\gamma$+jets events) 
can be used to study the background in the
hadronic $Z\gamma$ signal region. 

Tests to check whether the degrees of freedom of the chosen function 
are sufficient to accurately describe the background distribution in data 
are performed by comparing the goodness of the fits to the data 
using either the nominal background function or a function with 
one or two additional degrees of freedom.
A test statistic $\Lambda_{12}$ to discriminate between two
background models $f_1$ and $f_2$ is built.
This uses either the $\chi^2$ and number of degrees of freedom
computed from a binned comparison between the data and the fit
(leptonic analysis) or directly the maximum value of the likelihood
(hadronic analysis), for the fits performed to data using either $f_1$ or $f_2$.
The simpler model $f_1$ is then rejected in favour of $f_2$ if
the probability of finding values of $\Lambda_{12}$ more extreme than the one
measured in data is lower than 5\%.
No significant improvement in goodness of fit over the model of
Eq.~(\ref{back:eqn:background}) is found when adding one or two
extra degrees of freedom to it.

\section{Systematic uncertainties}
\label{sec:systematics}

The systematic uncertainty in the measured
$\sigma(pp\to X)\times BR(X\to Z\gamma)$
has contributions from uncertainties in the integrated
luminosity $L_{\mathrm{int}}$ of the analyzed data,
in the estimated signal yield $N_\mathrm{sig}$, and in the signal
efficiency $\varepsilon$. 

An integrated-luminosity uncertainty of $\pm 5\%$
is derived, following a methodology similar to that detailed in
Ref.~\cite{atlas:lumi2011}, from a preliminary calibration 
using $x$--$y$ beam-separation scans performed in August 2015. 

The uncertainties in the signal yield arise from the choice of
functional forms used to describe the signal and the background
in the final fit to \mrec,
as well as from the parameters of the
signal model, which are determined from the simulation.
Uncertainties due to the parameterization of the
signal distribution chosen in Section~\ref{ssec:signal_model} are negligible
compared to the other uncertainties.
Effects of spurious signals from the choice
of background function on the signal are included as described in
Section~\ref{ssec:background_model}.
The uncertainties in the signal model parameters
arise from the uncertainties in the energy scales
and resolutions of the final-state particles (photons, electrons,
muons, and large-radius jets).

Contributions to the uncertainty in the signal detection efficiency
$\varepsilon$
originate from the trigger and the reconstruction, identification and
isolation requirements of the selected final-state particles. There
is also a contribution from the kinematic requirements used to select
the final-state particles due to uncertainties in the energy scale and
resolution. 
The effects of the lepton and photon trigger, reconstruction, identification
and isolation efficiency uncertainties are estimated by varying the
simulation-to-data
efficiency correction factors by their $\pm 1\sigma$ uncertainties and
recalculating the signal efficiency.
The impact of the lepton and photon energy scale and resolution
uncertainties is estimated by computing the relative change in efficiency
and in the peak position and the width of the invariant mass distribution of
the signal after varying these quantities by their uncertainties
in the simulation.

The uncertainties in the jet $\pT$, mass and $D_2^{\beta=1}$
scales and resolutions are evaluated by comparing the ratio
of calorimeter-based to track-based measurements in dijet data
and simulation~\cite{ATL-PHYS-PUB-2015-033,ATLAS-CONF-2015-035}.
Their effect is estimated by recomputing the efficiency
of the hadronic $Z$ boson selection and the signal $m_{J\gamma}$
distribution after varying the $\pT$, mass and $D_2^{\beta=1}$
scales and resolutions by their uncertainties.
The requirement on the number of primary-vertex tracks associated
with the jet induces a 6\% systematic uncertainty in the
corresponding efficiency, as estimated from the comparison
of simulation and control samples of data. 

In the leptonic analysis, the systematic uncertainties have a
small effect on the final results, which are dominated by the 
statistical uncertainties originating from the small size of the selected
sample. The main contributions
arise from the uncertainty in the photon and electron resolution,
from the spurious signal and from the luminosity uncertainty. They
worsen the search sensitivity by only 4.0\%--0.5\%, 3.0\%--2.0\% and 0.5\%
respectively, over the $m_X$ range from 250~\GeV\ to 1.5~\TeV.

In the hadronic analysis, the systematic uncertainties
are dominated by estimates of the jet mass resolution and
the jet energy resolution. 
The search sensitivity worsens by 4.3\% (5.3\%),
4.3\% (1.1\%) and 2.1\% (1.0\%) at $m_{J\gamma}$ masses of 0.7~\TeV, 1.5~\TeV\
and 2.7~\TeV, from the effects of the jet mass resolution (jet energy resolution) 
uncertainty.
The degradation of the search sensitivity due to the uncertainty
in the efficiency of the requirement on the number of tracks
associated with the large-radius jet is less than 1\% at all tested masses.

\section{Statistical procedure}
\label{sec:interpretation}

A profile-likelihood-ratio method~\cite{stat} is used to search for
a localized excess over a smoothly falling background in the \mrec\ distribution
of the data, as well as to quantify its significance and estimate its
production cross section.
The extended likelihood function $\mathcal{L}(\alpha, \vecth)$ is given by the
product of a Poisson term, the values of the probability density function $f_\mathrm{tot}(\mrec^i,\alpha,\vecth)$ of the invariant mass distribution
for each candidate event $i$ and constraint terms $G(\vecth)$:
\begin{equation}
\LL\left((\alpha, \vecth)\Big|\{\mrec^i\}_{i=1..n}\right) = \frac{\mathrm{e}^{-N(\alpha, \vecth)}N^n(\alpha, \vecth)}{n!} \prod\limits_{i=1}^nf_\mathrm{tot}(\mrec^i,\alpha,\vecth) \times G(\vecth).
\end{equation}
In this expression $\alpha$ represents the parameter of interest, $\alpha=\sigma(pp\rightarrow X)\times BR(X\rightarrow Z\gamma)$, 
  $\vecth$ are {\em nuisance parameters},
$n$ is the observed number of events, and the expected event yield $N$ is the sum of the number of signal events $N_{\mathrm{sig}}=L_\mathrm{int}\times(\sigma\times BR)\times \varepsilon$, the number of background events $N_{\mathrm{bkg}}$, and  the spurious signal yield $N_\mathrm{spur}$ described in Section~\ref{ssec:background_model}.
The function $f_\mathrm{tot}(\mrec^i,\alpha,\vecth)$ is built from the signal and background probability density functions of \mrec, $f_\mathrm{sig}$ and $f_\mathrm{bkg}$:
\begin{multline}
  f_\mathrm{tot}(\mrec^i,\alpha,\vecth) = \frac{1}{N} \left[ \left( N_{\mathrm{sig}}(m_X, \alpha, \vecth_\mathrm{sig}) + N_{\mathrm{spur}}(m_X)\times \thetaspur \right) \right.
  \\ \left. \times f_{\mathrm{sig}}(\mrec^i,\vecth_\mathrm{sig}) + N_{\mathrm{bkg}} \times f_{\mathrm{bkg}}(\mrec^i,\vecth_\mathrm{bkg})\right].
\end{multline}
The uncertainties in the signal parameterization, efficiency and
bias in the signal yield due to the choice of the background model 
are included in the fit via nuisance parameters
which are constrained with Gaussian or log-normal penalty terms for
signal modelling and a Gaussian penalty term for the spurious signal
uncertainty.

The significance of the signal is estimated by computing
the $p$-value of the compatibility of the data with
the background-only hypothesis ($p_0$).
A modified frequentist ($CL_{s}$) method~\cite{cls} is used to set
upper limits on the signal cross section times branching ratio at 95\%
confidence level (CL), by identifying the value of $\sigma\times BR$ 
for which $CL_s$ is equal to $0.05$.

Closed-form asymptotic formulae~\cite{stat} are used to derive
the results.
Due to the small size of the selected dataset and of the expected
background for large values of $m_X$, the results for some
values of $m_X$, spread over the full tested range, are
checked using ensemble tests.
The results obtained using the
asymptotic formulae are in good agreement (differences on the
cross-section limits $<10\%$) with those from the ensemble tests
for most of the $m_X$ range, except at high $m_X$ where the differences
on the cross-section limits can be as large as 30\%.

\section{Results}
\label{sec:results}

In the data, there are 382 $Z(\to \ell\ell)\gamma$ candidates with 
$\mrec>200$~\GeV\ and 534 $Z(\to J)\gamma$ candidates with 
$\mrec>640$~\GeV.
The candidates with largest invariant mass in the leptonic and hadronic analyses
have $m_{\ell\ell\gamma}=1.47$~\TeV\ and $m_{J\gamma}=2.58$~\TeV\, respectively.

The invariant mass distributions of the selected $Z\gamma$ candidates in data
in the leptonic and hadronic final states are shown in
Figure~\ref{fig:min_data_bkgonly_fit}.
The solid lines represent the results of a background-only fit.

 \begin{figure}[!h]
  \begin{center}
    \subfigure[]{\includegraphics[width=0.7\columnwidth]{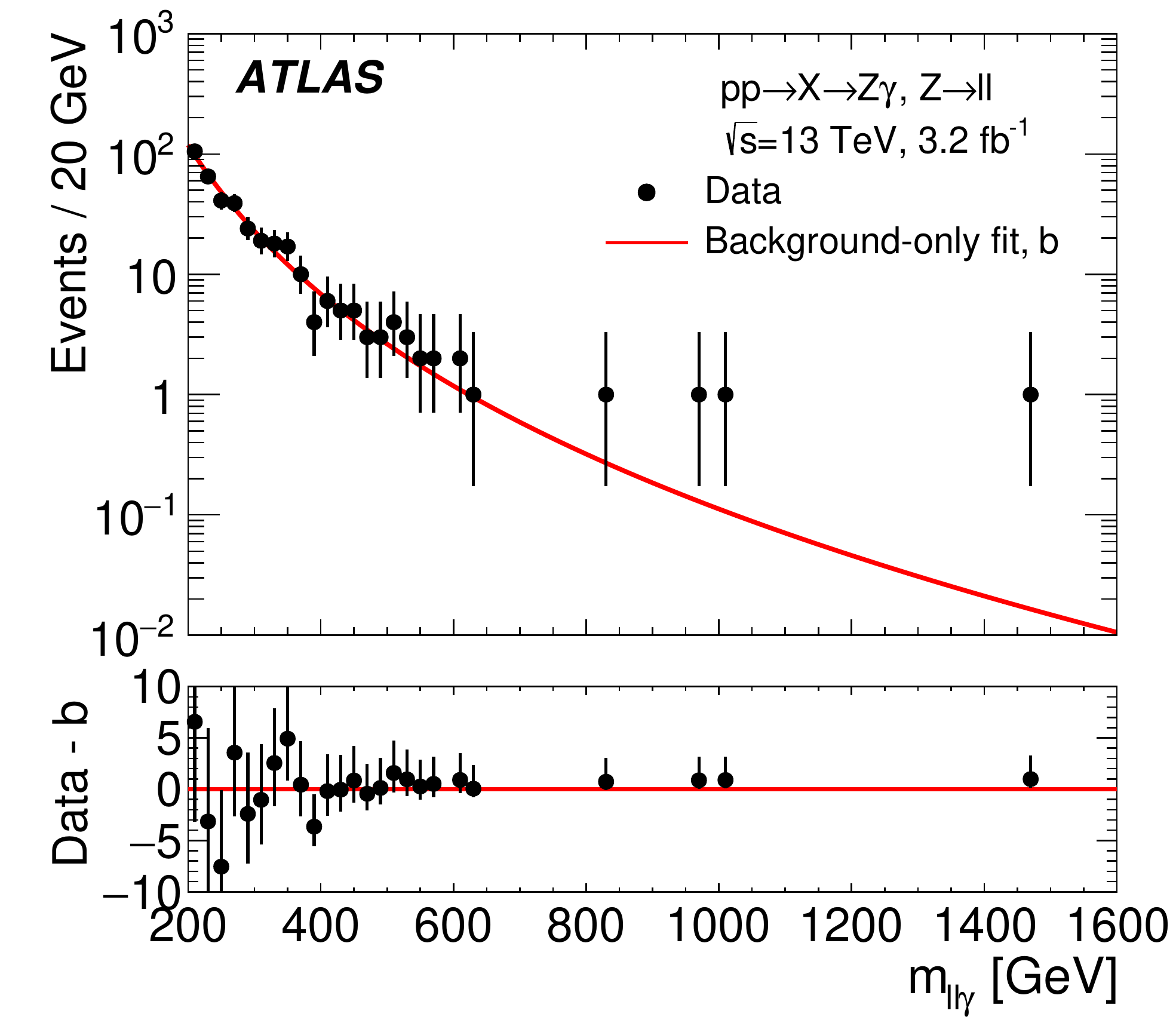}}
    \subfigure[]{\includegraphics[width=0.7\columnwidth]{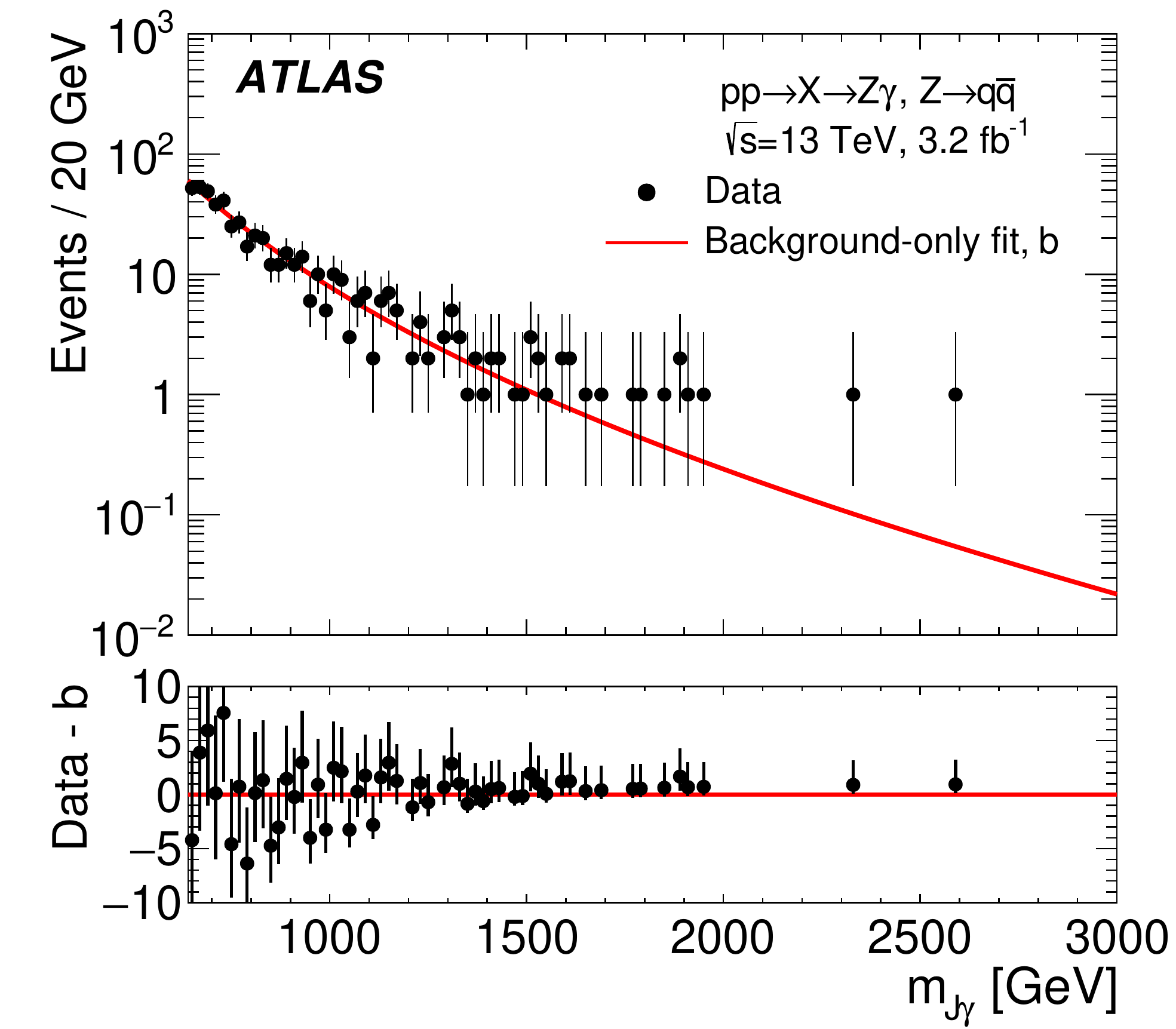}}
    \caption{Distribution of the reconstructed $Z\gamma$ invariant mass
      in events in which the $Z$ boson decays to (a) electron or muon pairs,
      or (b) to hadrons reconstructed as a single, large-radius jet.
      The solid lines show the results of background-only fits to the data.
      The residuals of the data points with respect to the fit are also shown.
    }
    \label{fig:min_data_bkgonly_fit}
  \end{center}
\end{figure}

There is no significant excess with respect to the background-only hypothesis,
and the largest deviations are observed 
around $m_X=350~\GeV$ in the leptonic analysis ($2.0\sigma$ local significance)
and around $m_X=1.9$~\TeV\ in the hadronic analysis ($1.8\sigma$
local significance).

For a narrow scalar boson $X$ of mass $m_X$, 95\% CL upper limits on
$\sigma(pp\to X)\times BR(X\to Z\gamma)$ are set
for $m_X$ between 250~\GeV\ and 1.5~\TeV\ in the leptonic analysis
and between 700~\GeV\ and 2.75~\TeV\ in the hadronic analysis.
In the $m_X$ range between 700~\GeV\ and 
1.5~\TeV\ the results of the two analyses are then combined.
The observed limits range between 295~fb for $m_X=340$~\GeV\ and 8.2~fb for
$m_X=2.15$~\TeV,
while the expected limits range between 230~fb for $m_X=250$~\GeV\
and 10~fb for $m_X=2.75$~\TeV. 
The observed and expected limits as a function of $m_X$ are shown in Figure~\ref{fig:limit_data_comb}.

\begin{figure}[!h]
  \centering
  \includegraphics[width=1.0\columnwidth]{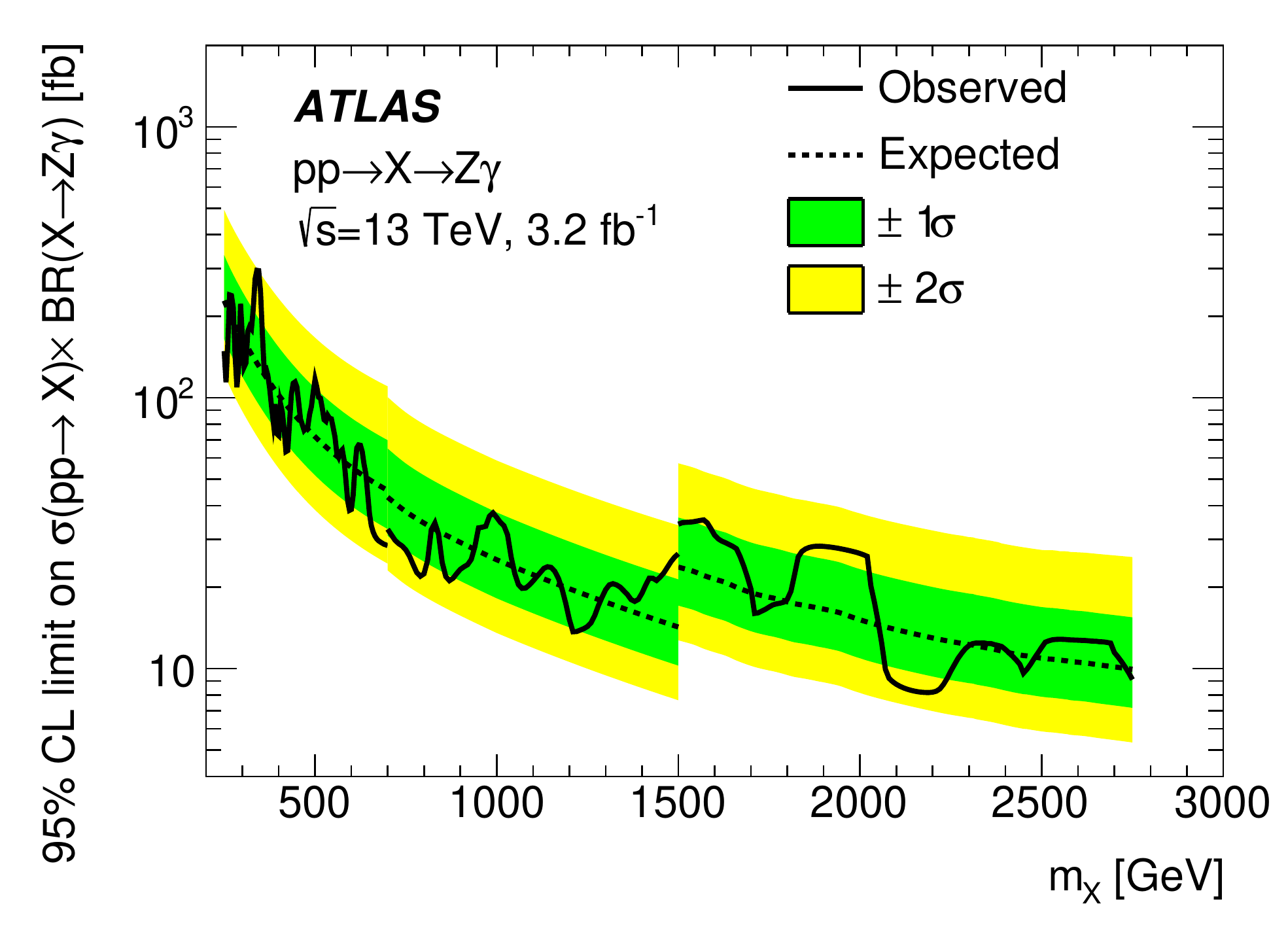}
  \caption{Observed (solid lines) and median expected (dashed lines)
    95\% CL limits on the product of the production cross section times
    the branching ratio of a narrow scalar boson $X$ decaying to a $Z$
    boson and a photon, $\sigma(pp\to X)\times BR(X\to Z\gamma)$,
    as a function of the boson mass $m_X$.
    The green and yellow solid bands correspond to the
    $\pm 1\sigma$ and $\pm 2\sigma$ intervals for the expected 
    upper limit respectively.
    The limits in the $m_X$ ranges of 250--700~\GeV\ and 1.5--2.75~\TeV\ are obtained from 
    the leptonic and hadronic analyses respectively, while in the range
    700~\GeV--1.5~\TeV\ they are obtained from the combination of the two analyses.
  }
  \label{fig:limit_data_comb}
\end{figure}

\FloatBarrier

\section{Conclusion}
\label{sec:conclusion}

A search for new resonances with masses between 250~\GeV\ and 2.75~\TeV\
decaying to a photon and a $Z$ boson
has been performed using 3.2~fb$^{-1}$ of proton--proton collision data
at a centre-of-mass energy of $\sqrt{s} = 13$~\TeV\
collected by the ATLAS detector at the Large Hadron Collider.
The $Z$ bosons were reconstructed through their decays either to charged,
light, lepton pairs ($e^+e^-$, $\mu^+\mu^-$) or to boosted quark--antiquark
pairs giving rise to a single, large-radius, heavy jet of hadrons.

No significant excess in the invariant-mass distribution of
the final-state particles due to a scalar boson with a narrow width 
(4~\MeV) was found over the smoothly falling background.

Limits at 95\%~CL using a profile-likelihood ratio method were set on
the production cross section times decay branching ratio to $Z\gamma$
of such a boson.
The observed limits range between 295~fb for $m_X=340$~\GeV\ and 8.2~fb for
$m_X=2.15$~\TeV,
while the expected limits range between 230~fb for $m_X=250$~\GeV\
and 10~fb for $m_X=2.75$~\TeV.

\section*{Acknowledgements}

We thank CERN for the very successful operation of the LHC, as well as the
support staff from our institutions without whom ATLAS could not be
operated efficiently.

We acknowledge the support of ANPCyT, Argentina; YerPhI, Armenia; ARC, Australia; BMWFW and FWF, Austria; ANAS, Azerbaijan; SSTC, Belarus; CNPq and FAPESP, Brazil; NSERC, NRC and CFI, Canada; CERN; CONICYT, Chile; CAS, MOST and NSFC, China; COLCIENCIAS, Colombia; MSMT CR, MPO CR and VSC CR, Czech Republic; DNRF and DNSRC, Denmark; IN2P3-CNRS, CEA-DSM/IRFU, France; GNSF, Georgia; BMBF, HGF, and MPG, Germany; GSRT, Greece; RGC, Hong Kong SAR, China; ISF, I-CORE and Benoziyo Center, Israel; INFN, Italy; MEXT and JSPS, Japan; CNRST, Morocco; FOM and NWO, Netherlands; RCN, Norway; MNiSW and NCN, Poland; FCT, Portugal; MNE/IFA, Romania; MES of Russia and NRC KI, Russian Federation; JINR; MESTD, Serbia; MSSR, Slovakia; ARRS and MIZ\v{S}, Slovenia; DST/NRF, South Africa; MINECO, Spain; SRC and Wallenberg Foundation, Sweden; SERI, SNSF and Cantons of Bern and Geneva, Switzerland; MOST, Taiwan; TAEK, Turkey; STFC, United Kingdom; DOE and NSF, United States of America. In addition, individual groups and members have received support from BCKDF, the Canada Council, CANARIE, CRC, Compute Canada, FQRNT, and the Ontario Innovation Trust, Canada; EPLANET, ERC, FP7, Horizon 2020 and Marie Sk{\l}odowska-Curie Actions, European Union; Investissements d'Avenir Labex and Idex, ANR, R{\'e}gion Auvergne and Fondation Partager le Savoir, France; DFG and AvH Foundation, Germany; Herakleitos, Thales and Aristeia programmes co-financed by EU-ESF and the Greek NSRF; BSF, GIF and Minerva, Israel; BRF, Norway; Generalitat de Catalunya, Generalitat Valenciana, Spain; the Royal Society and Leverhulme Trust, United Kingdom.

The crucial computing support from all WLCG partners is acknowledged gratefully, in particular from CERN, the ATLAS Tier-1 facilities at TRIUMF (Canada), NDGF (Denmark, Norway, Sweden), CC-IN2P3 (France), KIT/GridKA (Germany), INFN-CNAF (Italy), NL-T1 (Netherlands), PIC (Spain), ASGC (Taiwan), RAL (UK) and BNL (USA), the Tier-2 facilities worldwide and large non-WLCG resource providers. Major contributors of computing resources are listed in Ref.~\cite{ATL-GEN-PUB-2016-002}.

\printbibliography

\clearpage

\begin{flushleft}
{\Large The ATLAS Collaboration}

\bigskip

M.~Aaboud$^{\rm 135d}$,
G.~Aad$^{\rm 86}$,
B.~Abbott$^{\rm 113}$,
J.~Abdallah$^{\rm 8}$,
O.~Abdinov$^{\rm 12}$,
B.~Abeloos$^{\rm 117}$,
R.~Aben$^{\rm 107}$,
O.S.~AbouZeid$^{\rm 137}$,
N.L.~Abraham$^{\rm 149}$,
H.~Abramowicz$^{\rm 153}$
% ...
\bigskip
\\
% ...
$^{8}$ Department of Physics, The University of Texas at Arlington, Arlington TX, United States of America\\
% ...
$^{12}$ Institute of Physics, Azerbaijan Academy of Sciences, Baku, Azerbaijan\\
% ...
$^{86}$ CPPM, Aix-Marseille Universit{\'e} and CNRS/IN2P3, Marseille, France\\
% ...
$^{107}$ Nikhef National Institute for Subatomic Physics and University of Amsterdam, Amsterdam, Netherlands\\
% ...
$^{113}$ Homer L. Dodge Department of Physics and Astronomy, University of Oklahoma, Norman OK, United States of America\\
% ...
$^{117}$ LAL, Univ. Paris-Sud, CNRS/IN2P3, Universit{\'e} Paris-Saclay, Orsay, France\\
% ...
$^{135}$ $^{(a)}$ Facult{\'e} des Sciences Ain Chock, R{\'e}seau Universitaire de Physique des Hautes Energies - Universit{\'e} Hassan II, Casablanca; $^{(b)}$ Centre National de l'Energie des Sciences Techniques Nucleaires, Rabat; $^{(c)}$ Facult{\'e} des Sciences Semlalia, Universit{\'e} Cadi Ayyad, LPHEA-Marrakech; $^{(d)}$ Facult{\'e} des Sciences, Universit{\'e} Mohamed Premier and LPTPM, Oujda; $^{(e)}$ Facult{\'e} des sciences, Universit{\'e} Mohammed V, Rabat, Morocco\\
% ...
$^{137}$ Santa Cruz Institute for Particle Physics, University of California Santa Cruz, Santa Cruz CA, United States of America\\
% ...
$^{149}$ Department of Physics and Astronomy, University of Sussex, Brighton, United Kingdom\\
% ...
$^{153}$ Raymond and Beverly Sackler School of Physics and Astronomy, Tel Aviv University, Tel Aviv, Israel\\
% ...
$^{a}$ Also at Department of Physics, King's College London, London, United Kingdom\\
$^{b}$ Also at Institute of Physics, Azerbaijan Academy of Sciences, Baku, Azerbaijan\\
$^{c}$ Also at Novosibirsk State University, Novosibirsk, Russia\\
$^{d}$ Also at TRIUMF, Vancouver BC, Canada\\
$^{e}$ Also at Department of Physics {\&} Astronomy, University of Louisville, Louisville, KY, United States of America\\
% ...
\end{flushleft}


\end{document}